\documentstyle[epsfig,a4,12pt]{article}

\begin{document}
%

\newcommand{\Author}    {\large\bf Umut G\"ursoy$^{a,b}$ and Cihan Sa\c cl\i o\~glu$^{c,b}$}

\def\toprule{\noalign{\hrule \medskip}}
\def\midrule{\noalign{\medskip\hrule }}
\def\botrule{\noalign{\medskip\hrule }}
\setlength{\parskip}{\medskipamount}
\newcommand{\ud}{{\mathrm{d}}}


\begin{titlepage}
%
%
%
%
\begin{center}
    \huge\bf\boldmath
An open bosonic string with one end fixed
\end{center}
\bigskip
%
%
\begin{center}
\Author
\end{center}

\begin{center}
{\small
$^a$ Center for Theoretical Physics, MIT, 02139 Cambridge, Massachusetts, USA\\
$^b$ Feza G\"ursey Institute, TUBITAK--Bo\~gazi\c ci University, 81220, Istanbul, Turkey\\
$^c$ Physics Department, Bo\~gazi\c ci University, 80815 Bebek, Istanbul, Turkey }
\end{center}

\bigskip
%
%
\begin{abstract}
We study a bosonic string with one end free and the other confined
to a $D0-$brane.  Only the odd oscillator modes are allowed, which
leads to a Virasoro algebra of even Virasoro modes only.  The
theory is quantized in a gauge where world-sheet time and ordinary
time are identified.  There are no negative or null norm states,
and no tachyon.  The Regge slope is twice that of the open string;
this can serve as a test of the usefulness of the the model as a
semi-quantitative description of mesons with one light and one
extremely heavy quark when such higher spin mesons are found. The
Virasoro conditions select specific $SO(D-1)$ irreps. The
asymptotic density of states can be estimated by adapting the
Hardy-Ramanujan analysis to a partition of odd integers; the
estimate becomes exact as $D$ goes to infinity.

\end{abstract}

\end{titlepage}

\pagenumbering{arabic}

\section{Introduction}

Before its reincarnation as a candidate for a fundamental Theory
of Everything, String Theory first entered Physics
\cite{Nambu},\cite{Susskind} as an attempt to understand the
spectrum and interactions of hadrons. This attempt enjoyed
considerable success in at least qualitatively accounting for many
experimentally observed features of Strong Interactions.  For
example, the Veneziano amplitude \cite{Venez} exhibited
Dolen-Horn-Schmid duality \cite{DolenHS}, appropriate high $s$ and
high $t$ behaviors, and linear Regge trajectories.  The predictive
power of the model was further increased by combining strings with
the Quark Model.  In this picture, a meson was represented as a
quark-antiquark pair connected by a flux tube which behaved like a
string.  It thus became possible to place observed hadrons in
leading or daughter Regge trajectories and predict the masses,
spins and internal quantum numbers of mesons that were yet to be
observed. In a similar way, Harari-Rosner
\cite{Harari},\cite{Rosner} diagrams provided a useful way of
explaining many qualitative properties of particular hadronic
processes. This line of research was abandoned upon the almost
simultaneous realization that the microscopic theory of Strong
Interactions had to be based on an unbroken Yang-Mills theory and
that String dynamics were really tailored for 10 or 26 dimensions.

In this note we would like take a modest backward step in history and
present a qualitative
model for the excited states of mesons consisting of one very
heavy quark/antiquark (of the Charmed,
Bottom or the Top kind) and one very light antiquark/quark (Up or Down) in
terms of a string with one end fixed and one end free.  In String Theory language,
this system may be approximately described by using Dirichlet boundary conditions
(placing the infinitely heavy
quark at the origin of the space coordinate system, for example) at one end and
Neumann boundary conditions at the other. The resulting system, which we will refer
to as a Neumann-Dirichlet, or ND string, is worth examining
not only as a guide to some expected properties of mesons with one very heavy
quark or antiquark, but also for its novel String theoretic aspects. In fact, our
emphasis will be more on the latter question, given that the data on Regge
recurrences of such mesons is not available at this stage.

The plan of the paper and some of our main results are as follows.
In section 2, we apply the ND conditions on the string in a
natural adaptation of the light cone gauge of the NN string to our
problem.  This leads to what might be called a "rest frame gauge".
The ND conditions only allow odd oscillator modes, which are then
canonically quantized. The spacetime Hamiltonian and the
world-sheet energy momentum tensor are given. In section 3, the
Virasoro generators and their algebra for the ND string are worked
out.  The use of only odd oscillators implies that only even
Virasoro mode operators survive and the central term is half that
of the usual bosonic string. In section 4, the normal ordering
constant for $L_0$ is calculated and found to have an opposite
sign relative to the usual one; consequently, the ND string has no
tachyon and no critical spacetime dimension. The absence of even
oscillator modes and the tachyon have also been noted earlier in
\cite{Kleinert}. As excited states are now obtained using odd
creation operators only, the spectrum is a subset of that of the
NN string.  We then present a simple classical model to check the
mass-angular momentum relationship for the leading Regge
trajectory. The asymptotic density of states can be calculated
using a modified version of the Hardy-Ramanujan analysis for the
open string. Section 5 ends the paper with concluding remarks.

\section{Quantization}

We start with the Bosonic string action in a $D$-dimensional Minkowski
spacetime, hence

\begin{equation}
S=-\frac{T}{2} \int \ud \sigma \ud \tau
g^{\alpha\beta} \partial_{\alpha} X^{\mu}\partial_{\beta}X^{\nu}
\eta_{\mu\nu}.\end{equation}

We fix the reparametrization and Weyl invariances of the action
in the conformal gauge $g^{\alpha\beta}=\eta^{\alpha\beta}$.
The general solutions of the equations of motion

$$ (\partial^2_{\sigma}- \partial^2_{\tau})X^{\mu}=0$$
can be Fourier expanded as

\begin{equation}\label{gensol}
X^{\mu}=x^{\mu}+l^2p^{\mu}\tau+\frac{l}{2}i\sum_{n\ne 0}
\frac{\alpha^{\mu}_{n}}{n}\exp{-in(\tau-\sigma)}+
\frac{l}{2}i\sum_{n\ne 0}\frac{\tilde{\alpha}^{\mu}_{n}}{n}\exp{-in(\tau+\sigma)},
\end{equation}
where $l$ is a fundamental length introduced to ensure the
correct dimensions for $X^{\mu}$.
It is related to the string tension $T$ by $l=\sqrt{\frac{2}{\pi T}}$  .
The range of the parameter $\sigma$ is taken to be $[0,\pi/2]$ for reasons
which will soon be clear.
We locate the infinitely massive quark at the $\sigma=0$ end and also
identify this point with the origin of space coordinates through the Dirichlet
boundary condition

\begin{equation}\label{dirich}
X^{i}(0,\tau)=0,
\end{equation}
which gives $x^{i}=p^i=0$ and $\alpha_n^i=-\tilde{\alpha}_n^{i}$
for $i=1,2,..,D-1$. In modern parlance, we confine one end of the
string to a $D0$-brane \cite{Polch}. We will also take $x^{0}=0$.
We adopt the Neumann boundary conditions

\begin{equation}\label{neumann}
\partial_{\sigma}X^{\mu}(\pi/2,\tau)=0
\end{equation}
for the massless end, which require that $\alpha_n^{\mu}=0$ for even $n$.
This resembles the vanishing of the even parity modes in the
familiar ordinary quantum mechanics problem of a
particle placed in a "half-oscillator potential",
i.e., a well with an infinite barrier at $x=0$ and a harmonic oscillator
potential for $x>0$.

Imposing the ND boundary conditions on the space components thus
results in

\begin{equation}\label{gensoli}
X^{i}=-l\sum_{k}\frac{\alpha^{i}_{2k+1}}{2k+1}\exp{(-i(2k+1)\tau)}\sin{((2k+1)\sigma)},
\end{equation}
with the reality condition yielding $\alpha_n^{i\dagger}=\alpha_{-n}^i$ as usual.

We now turn to $X_0$.  Here, an analogy with the light-cone treatment of the
NN string suggests itself.  In the light-cone gauge,
the $\tau$-parameter is identified with one of the light-cone target
coordinates via

\begin{equation}\label{light--cone}
X^+(\sigma, \tau)= x^{+} +p^{+}\tau ,
\end{equation}
and this sets the $\alpha^{+}_{n}$ operators to zero. Using the constraint equations,
one then solves for the $\alpha^{-}_{n}$ in terms of the $D-2$ independent
transverse operators; hence there is no further need for imposing Virasoro gauge
conditions on the states.  By contrast, in old covariant quantization, all $X_{\mu}$
are treated on an equal footing, but Virasoro conditions are imposed on the states.

In our case, only one end of the string describes a lightlike worldline,
while the other is timelike; hence, the full light--cone gauge treatment
is inappropriate. Instead,
the natural analog of (\ref{light--cone}) is to identify the target space and
world sheet "times" via

\begin{equation}\label{gauge}
X^{0}=l^2p^0\tau.
\end{equation}

This eliminates the $\alpha^{0}_{n}$ modes and
consequently, states with negative norm.
We thus adopt an intermediate approach between light--cone
and old covariant quantization.
There will still be a need to impose constraint equations on the states, although
these have been partially pruned by (\ref{gauge}).
This will lead to a Virasoro algebra with special features which we can
already anticipate: By fixing one end of the string in target space,
we have broken Poincar\'{e} invariance down to $SO(D-1)$ rotational
symmetry about the origin of space.  Just as in a mechanical problem with an
infinite mass, we no longer expect momentum conservation to hold, while energy
should still be conserved.  This leads us to expect the $L_1$ and $L_{-1}$ modes,
which involve the momentum operator, will be absent;
indeed, since any odd mode can generate these unallowed modes through
commutation, only even Virasoro modes should be present.  We will see that this
actually happens.

We follow the conventional canonical quantization procedure by requiring that the
canonical momenta $\Pi$ and and the coordinates obey the equal time Poisson brackets

$$\{\Pi^i(\sigma),X^j(\sigma')\}=-\delta^{ij}\delta(\sigma-\sigma').$$
Insertion of (\ref{gensoli}) into this and the use of the formula
$$\pi\delta_{2\pi}(\sigma-\sigma')=\sum_{k}\exp(i(2k+1)(\sigma-\sigma'))$$
which may be obtained from the Poisson summation formula
(a factor of $\cos \frac{(\sigma-\sigma')}{2}$ in the denominator of the
LHS is omitted since it becomes unity when the argument
of the delta function vanishes)
yield the expected Poisson brackets ,
$$\{\alpha_n^i,\alpha_m^j\}=in\delta_{n,-m}\delta^{ij}$$
for the Fourier components.

Quantization is effected by replacing Poisson brackets with commutators with
the result
\begin{equation}\label{comrel}
[\alpha_n^i,\alpha_m^j]=n\delta_{n,-m}\delta^{ij}.
\end{equation}
This is a subalgebra of the bosonic string Heisenberg algebra, $n,m$ being
restricted to odd integers.

The Hamiltonian of the theory reads
\begin{eqnarray}\label{ham}
H=\frac{T}{2}\int_0^{\pi/2}(:\partial_{\tau}X^{\mu}
\partial_{\tau}X_{\mu}:+:\partial_{\sigma}X^{\mu}
\partial_{\sigma}X_{\mu}:)\ud\sigma\nonumber\\
=\frac{T}{2}\int_0^{\pi/2}(:\partial_{\tau}X^{i}
\partial_{\tau}X^{i}:+:\partial_{\sigma}X^{i}
\partial_{\sigma}X^{i}:-l^4p_0^2)\ud\sigma
\end{eqnarray}
with the gauge choice (\ref{gauge}). The infinite ground state energy caused
by the normal ordering procedure will be calculated and regularized later.
The energy--momentum tensor

$$T_{\alpha\beta}=\frac{T}{2}(:\partial_{\alpha}X^{\mu}
\partial_{\beta}X_{\mu}-\frac{1}{2}\eta_{\alpha\beta}
\partial_{\gamma}X^{\mu}\partial^{\gamma}X_{\mu}:)$$
has the components
\begin{eqnarray}\label{emt}
T_{00} & = & T_{11}=\frac{T}{4}(:\partial_{\tau}X^{i}
\partial_{\tau}X^{i}:+:\partial_{\sigma}X^{i}\partial_{\sigma}X^{i}:-l^4p_0^2),\\
T_{01} & = & T_{10}=\frac{T}{2}:\partial_{\tau}X^{i}\partial_{\sigma}X^{i}:.
\end{eqnarray}

\section{An even Virasoro algebra}

We now verify our earlier claim that the ND string leads to a
Virasoro algebra of even modes only.  Imposing open string boundary
conditions of the usual NN type
identifies the oscillators for the left-moving modes and the right-moving modes,
leading to a single set of Virasoro operators.  In the ND string, we not only
identify left and right-moving oscillators, but in addition exclude the even modes.
It is therefore not surprising that half of the Virasoro modes
will be eliminated. These have to be the odd ones which cannot close upon
commutation, while the even Virasoro modes obviously can.  In order to compute
them, we first observe that

\begin{equation}\label{T++}
T_{\pm \pm} = \frac{1}{2}(T_{00} \pm T_{01})
= \frac{T}{8}(:(\partial_{\tau}X^{i} \pm \partial_{\sigma}X^{i})^{2}:
 - (l^{2}p^{0})^{2}),
\end{equation}
where
\begin{equation}\label{Xpm}
\partial_{\tau}X^{i} \pm \partial_{\sigma}X^{i}=
\mp l \sum_{m=odd}\alpha_m^{i} \exp{(-in(\tau \pm  \sigma))}.
\end{equation}

Thus $T_{++}(\sigma)= T_{--}(-\sigma)$ as in the ordinary open string.  Using
this property and also the fact that for $\tau=0$

\begin{equation}
:(\partial_{\tau}X^{i} \pm \partial_{\sigma}X^{i})^{2}:=
l^{2} :\sum_{n=odd}\sum_{n=odd} \exp{(-im\sigma)} \alpha_n^{i}\alpha_{m-n}^{i}:,
\end{equation}
where $m$, being the sum of two odd integers, is now necessarily even, we
arrive at the even Virasoro mode operators

\begin{equation}
L_n = 2\int_{\frac{-\pi}{2}}^{\frac{\pi}{2}} \exp{in\sigma}\ T_{++} d\sigma.
\end{equation}

The constraints $T_{\alpha \beta} = 0$ can now be imposed by demanding
$L_n$ annihilate physical states for $n>0$ and $L_o = - \epsilon_C$ on the same states.
The number $\epsilon_C$ will be determined later.

Calling $\alpha_0^{\mu}=lp^{\mu}$, we can read off
\begin{eqnarray}\label{symgens}
L_0 & = & -\frac{1}{2}(\alpha_0^0)^2+\frac{1}{2}\sum_{m=odd}:
\alpha_m^{i}\alpha_{-m}^{i}:,\\
L_n & = &\frac{1}{2}\sum_{m=odd}:\alpha_m^{i}\alpha_{n-m}^{i}:
\qquad \textrm{(n even and non-zero)} \quad, \\
L_n & = &0 \qquad \textrm{ (n odd)}.
\end{eqnarray}

We note that the Virasoro operators have their usual form in terms of the
odd oscillator modes, hence it is clear that they will satisfy the familiar Virasoro
algebra except perhaps for a change in the central extension term.
This term is most easily calculated via the
vacuum expectation value

\begin{eqnarray}
A_m & = &\langle 0| \lbrack L_m,L_{-m}\rbrack |0\rangle
= \langle 0| L_m  L_{-m} |0\rangle\nonumber\\
& = &  \frac{1}{4}\langle 0|\sum_{n,k=odd}\alpha_{m-n}^{i}
\alpha_{n,i} \alpha_{-m-k}^{j}\alpha_{k,j} |0\rangle\\
& = &  \frac{1}{4}\langle 0|\sum_{n,k=odd}\alpha_{n}^{i}
\alpha_{m-n,i} \alpha_{k}^{j}\alpha_{-m-k,j}|0\rangle\nonumber,\\
\end{eqnarray}
where $m$ is of course even and positive.
The expectation value in the second line vanishes for $k>0$ and for $n>m$,
whereas the expectation value
in the third line vanishes for $n<0$ and for $k<-m$, yielding

\begin{eqnarray}
A_m & = &  \frac{1}{4}\langle 0|\sum_{n=1}^{m-1}\sum_{k=-m+1}^{-1}
\alpha_{m-n}^{i}\alpha_{n,i} \alpha_{-m-k}^{j}\alpha_{k,j}|0\rangle\nonumber\\
& = &  \frac{D-1}{2}\sum_{n=1}^{m-1}n(m-n)\\
\end{eqnarray}
where all the summations above are over odd integers.
The second line is obtained by successive use of (\ref{comrel}) and
$D-1 =\eta_{ij}\eta^{ij}$ is the dimension of space.
The summation in the second line, when taken over all integers, produces
the well--known Virasoro central extension term.  Our odd--integer summation,
in contrast, yields
a modified central extension for the even Virasoro algebra of the form

\begin{equation}\label{eVir}
\lbrack L_n,L_m \rbrack = (n-m)L_{m+n}+\frac{D-1}{24}\delta_{n+m,0}(n^2+2)n.
\end{equation}

It should be noted that the anomaly-free $SL(2,R)$ subalgebra of the open string Virasoro
algebra is now reduced to $U(1)$.
It is possible to recover a more familiar form for the central term
by shifting $L_0$ to $L_0 -\frac{D-1}{16}$.  This gives

\begin{equation}\label{eVir2}
\lbrack L_n,L_m \rbrack = (n-m)L_{m+n}+\frac{D-1}{24}\delta_{n+m,0}(n^3-n).
\end{equation}

Note that the coefficient $D$ in the usual Virasoro algebra
has been changed to $\frac{D-1 }{2}$.  The elimination of the odd Virasoro modes is
seen to manifest itself as an apparent halving of the number of bosonic coordinates.

\section{The spectrum}

The on--mass shell condition comes from the zero mode of the
Virasoro constraints $T_{\alpha\beta}=0$.
Specifically, we derive the masses of the
states at the quantum level from

$$ (L_0+\epsilon_C)|N, p\rangle=0,$$
where $\epsilon_C$ is the Casimir energy which emerges upon normal ordering of
the oscillators.  We note that given one end is fixed, all the excitations
will have to be analysed in their common rest frame. We thus combine

$$ m^2=-p^{\mu}p_{\mu}=-l^{-2}\alpha_0^{\mu}\alpha_{0,\mu}=(\alpha_0^0)^2,$$
with the zero mode Virasoro constraint

\begin{equation}\label{mass}
\frac{l^2m^2}{2}=\epsilon_C+N =
\epsilon_C + \sum_{i=1}^{D-1}
\sum_{m=1}^{\infty}\alpha^i_{-(2m-1)}\alpha^i_{2m-1}.
\end{equation}

In the above, we have performed the normal ordering.  The Casimir energy is
given by

\begin{equation}\label{casimir}
2 \epsilon_C=(D-1)R\left\{\sum_{m=1}^{\infty}(2m-1)\right\}.
\end{equation}

Here $R\{ .\}$ denotes the regularized part of the expression inside the brackets.
We regularize $\epsilon_C$ via

\begin{equation}\label{regul}
2 \epsilon_C=(D-1)\lim_{\lambda\to 0}
\sum_{m=1}^{\infty}(2m-1)\exp(-\lambda (2m-1)).
\end{equation}
where it is understood that the term that diverges as $\lambda$ goes to zero
will be discarded at the end.

Doing the summation, one finds that to second order in $\lambda$

\begin{equation}\label{regul2}
\frac{2}{D-1} \epsilon_C=
\lim_{\lambda\to 0}\frac{1}{\lambda^2}+\frac{1}{12}+O(\lambda^2),
\end{equation}
from which we read off the regularized Casimir energy as

\begin{equation}\label{casimir2}
\epsilon_C=\frac{D-1}{24}.
\end{equation}

A quicker way to arrive at the same result is to
note that $\zeta(-1)- 2\zeta(-1)= \frac{1}{12}$.  The mass formula thus becomes

\begin{equation}\label{mass2}
m^2=\frac{1}{l^2}(\frac{D-1}{12}+ 2N),
\end{equation}
where $N$ is given by

$$N=\sum_{i=1}^{D-1}\sum_{m=1}^{\infty}\alpha^i_{-(2m+1)}\alpha^i_{2m+1}.$$

Since $N$ also denotes the highest spin at a given mass, we see
that the ND string exhibits linear Regge trajectories,
albeit with a Regge slope

$$\alpha'_{ND}= \frac{1}{\pi T}$$
which is twice as large as the usual $\alpha'_{NN}$. The reason behind this
can be seen most clearly by comparing an NN string and an ND string
both rotating in their rigid, maximum spin modes, the former about its
geometrical center, the latter about the fixed end.
When data on the excited states of
mesons with one light and one extremely heavy quark/antiquark become available, it
would be interesting to see whether this slope doubling is experimentally confirmed.

The above conclusions about the Regge trajectories are supported by examining
an ND string undergoing rigid rotation in the $X^1 X^2-$plane with

\begin{equation}\label{classical}
X^1=l^2 p^0 \sin\sigma \cos\tau, X^2=l^2 p^0 \sin\sigma\sin\tau, X^0=l^2 p^0 \tau.
\end{equation}
It can readily be verified that these satisfy both the equations of motion and
the constraints of vanishing world sheet energy momentum tensor.

The energy of the configuration is

\begin{equation}\label{M}
E=M=T\int_{0}^{\frac{\pi}{2}}\partial_{0}X^{0} d\sigma=
\frac{l^2 p^0 \pi T}{2}
\end{equation}
while its angular momentum turns out to be
\begin{equation}\label{J}
J=T\int_{0}^{\frac{\pi}{2}}(X^1 \partial_0 X^2 - X^2\partial_0 X^1)
=\frac{\pi T (p^0 l^2)^2}{4}.
\end{equation}
The rotation being rigid, angular momentum per $M^2$ is maximised. The ratio of
the former to the latter is the Regge slope, which again comes out to be
$\frac{1}{T\pi}$.

We can draw some important conclusions from (\ref{mass2}).  First
of all, the vacuum state with no excitations $(N=0)$, is a
positive mass state with $M^2=\frac{D-1}{12 l^2}$, unlike the
tachyon in the NN bosonic string. Hence the model has a stable
vacuum.  The second conclusion concerns the critical dimension for
the ND string.  From what we have seen so far, it might at first
appear that the ND string can exist in any dimension $D>1$. To see
why this looks plausible, first recall that we eliminated the
negative probability modes at the beginning by identifying $X^0$
with $\tau$ and setting all the $\alpha^{0}_{n}$ equal to zero. As
the fixed end condition breaks $D$ dimensional Poincar\'{e}
invariance down to $SO(D-1)$, all we have to check now is whether
this reduced rotational symmetry holds in the theory independently
of the value of $D$.  To show this, we can take over Polchinski's
argument \cite{Polch'} used for deriving the critical dimension
for the bosonic NN string: By seeking consistency between the
Lorentz transformation properties of the states created by the
oscillator modes and the mass formula, $D=26$ emerges without
having to check the closure of the Lorentz algebra.  This amounts
to verifying Lorentz covariance in representation space instead of
on the operator algebra. In our case, all of the states come in
irreducible representations of $SO(D-1)$ since they are created by
the $D-1$ space components of the oscillators, and they are all
massive, as can be seen from the mass formula.  Thus $SO(D-1)$
symmetry is realized on the states regardless of the value of $D$;
in fact, as we will see below, the Virasoro conditions select
specific $SO(D-1)$ irreps as states.  However, the presence of an
anomaly term proportional to $D-1$ in our even Virasoro algebra
renders this conclusion suspect; we will return to this paradox in
the last section.

Turning to the spectrum, the first few examples are as follows:
\begin{itemize}
\item
$N=1$: $$|1\rangle=\alpha_{-1}^i|0\rangle$$
is a massive vector representation;
\item
$N=2$: $$|2\rangle=\alpha_{-1}^i\alpha_{-1}^j|0\rangle$$
can be decomposed into a symmetric traceless massive rank-2 $SO(D-1)$ tensor
and a massive scalar;
\item
$N=3$: $$|3\rangle=\alpha_{-3}^i|0\rangle\oplus\alpha_{-1}^l
\alpha_{-1}^j\alpha_{-1}^k|0\rangle$$
can be decomposed a symmetric traceless rank-3 tensor plus 2 vectors.
\end{itemize}

At this point one may wonder what role is left for the other Virasoro conditions
$L_{n>0}|phys\rangle = 0$ to play.  After all, negative-metric Hilbert space states have
already been eliminated at the outset, and linear combinations of the states
above can be used to build perfectly physical looking $SO(D-1)$ irreps.  The answer,
alluded to above, is that not all the
possible $SO(D-1)$ irreps at a given $N$ are allowed.  The reader can quickly
verify that while the $N=1$ state is automatically annihilated by $L_{2}, L_{4}, ...$,
at $N=2$, it is only the rank-2 tensor that satisfies the same Virasoro conditions.
Similarly, at $N=3$, the Virasoro conditions select the rank three tensor and the
vector created by $\alpha_{-3}^{i}$, while ruling out the vector
$\alpha_{-1}^{i}\alpha_{-1}^{k}\alpha_{-1}^{k}$, just as the massive scalar
$\alpha_{-1}^{k}\alpha_{-1}^{k}$ at $N=2$ is ruled out.
Note that the highest rank tensor state at a given $N$, built out of $N$ $\alpha^i_{-1}$
oscillators (with appropriate subtractions of other
lower irreps that are created along the way) always exists, and so does the vector
state obtained from $\alpha_{-N}^{i}$.
Also, one can still obtain states with even $N$ although only combinations of
odd-mode creation operators are being used.  Thus we can for example write
$4 = 1+1+1+1$ or $4=1+3$, but are not allowed to use $4=4, 4=2+1+1, 4=2+2$.
This means the asymptotic level density as $N\gg 1$ and the Hagedorn temperature of the
ND string will not be identical with those of the NN string.  We can compute these
quantities by slightly modifying the standard techniques given in, say, \cite{gsw3}.
One can derive the number of states $d_n$ for $N=n$ (or the
multiplicity at level $n$ of the partition function $tr w^N$) via

\begin{equation}\label{dense}
d_n=\frac{1}{2i\pi}\oint\ud w w^{-n-1}tr w^N.
\end{equation}

We thus need to compute

$$tr w^N=g(w)^{-(D-1)}$$
where $g(w)=\prod_{n=1,3,5,\dots}(1-w^n)$. There is actually a subtle point
here.  The original version of this calculation is based on an entirely
physical spectrum built
out of $D-2=24$ transverse oscillators in the light-cone gauge and therefore
the $d_n$ represent the true number of physical states at a given $n$.
Our calculation, in contrast, will yield the number of all of the states generated by
combinations of the $\alpha_{-m}^{i}$ before the Virasoro conditions eliminate the
unphysical ones.  The relative error in this overestimation however decreases
as $1/(D-1)^2$ with increasing $D$ since the unwanted states come from the
contraction of the space indices of two oscillators to give a scalar.

Now obviously

$$g(w)=\frac{f(w)}{f(w^2)},$$
where $f(w)=\prod_{n=1}^{\infty}(1-w^n)$ is related to the Dedekind eta function by

$$\eta(\tau)=\exp(i\pi\tau/12)f(\exp(2i\pi\tau)).$$
Here $w= \exp(2i\pi\tau)$ as usual.

Using the S-transformation property of the eta function

$$\eta(-1/\tau)=(-i\tau)^{1/2}\eta(\tau), $$
one gets

$$g(w)=\frac{1}{\sqrt{2}}w^{1/24}q^{1/24}\frac{f(q^2)}{f(q)},$$
where $q=\exp(\frac{2\pi^2}{\ln(w)})$.

To investigate the asymptotic density of states we look at the limit
$w\to 1$ for which

$$g(w)\to A\exp(\frac{\pi^2}{12\ln{w}}),$$
or

$$tr w^N\to A\exp(\frac{-\pi^2(D-1)}{12\ln{w}}),$$
where $A$ is a constant. Using (\ref{dense}) we have

\begin{equation}\label{dense2}
d_n\approx\frac{1}{2i\pi}\oint\ud w \exp(-\frac{\pi^2(D-1)}{12\ln{w}}-(n+1)\ln w).
\end{equation}

Performing the saddle point approximation in (\ref{dense2}) around the saddle point

$$\ln w^\star=-\pi\sqrt{(D-1)/12(n+1)}$$
one arrives at the
asymptotic expression for the level density,

\begin{equation}\label{dense3}
d_n\approx \exp(\pi \sqrt{\frac{(D-1)n}{3}})(D-1)^{1/4}n^{-3/4},
\end{equation}
where we have only displayed the dependence on the dimension of the space--time
and the level $n$, omitting some multiplicative constants. One can compare
this with the formula for the NN bosonic string in $D$ dimensions \cite{gsw4}:

\begin{equation}\label{dense4}
d_n\approx \exp(4\pi \sqrt{\frac{n(D-2)}{24}})n^{-3/4}n^{-(D-2)/4}.
\end{equation}

We observe that going from (\ref{dense3}) to (\ref{dense4}), the argument
of the exponential changes from $\sqrt{n}$ to $\sqrt{2n}$,
This stems from the fact that the ND spectrum is built using only
odd spin oscillators, whereas in the NN string
both odd and even oscillators contribute to create a state at level $n$.
Nevertheless, one can see that $d_n(ND)>d_n(NN)$ for $D=2$ and $D=3$, which
is due to the relatively larger contribution of the longitudinal oscillator modes
to the ND string for these low dimensions.  For higher $D$, the exponential dominates
as $n$ goes to infinity and the NN string has the greater
multiplicity at a given large $n$.

The level formula (\ref{dense3}) shows that the Hagedorn temperature of the ND
string is

\begin{equation}\label{haged}
T_{ND} = \frac{1}{\pi}\sqrt{\frac{3}{2\alpha'_{NN}(D-1)}}.
\end{equation}
Here we used the fact that $\alpha'_{ND}=2\alpha'_{NN}$.  The Hagedorn
temperature for the NN string in $D$ dimensions follows from the same
formula when $D-1$ is replaced by $D-2$.

\section{Concluding Remarks}

The ND string described above may have some qualitative relevance
for describing mesons with one very heavy quark and a very light
antiquark. For example, finding such mesons on Regge trajectories
with twice the slope of the light meson trajectories would be a
confirmation of the picture proposed here. However, we believe the
ND string is interesting enough to be considered on its own, just
as the original bosonic string proved to have a significance
beyond hadron phenomenology.  We have shown that it is a bona fide
solution of the string equations of motion obeying the constraints
and exhibiting a characteristic spectrum: it is essentially the
system obtained by viewing the open string from a frame centered
at its mid-point and then throwing half of the string away.  The
constraints lead to a physical Hilbert space via a combination of
modified Virasoro conditions and a choice of the $\tau$ parameter
which eliminates negative probabilities. However, there are three
questions that need to be answered; the first two turn out to be
related.  The first is the unaddressed expectation that the
problem should involve an infinite mass associated with the fixed
end. The second and the third involve the apparent conflict
between two facts: On the one hand, the theory appears to be
unitary and $SO(D-1)$ invariant for any $D$, while on the other
hand, the presence of a non-zero anomaly term in the Virasoro
algebra casts doubt on the possibility of imposing the constraints
$T_{\alpha \beta}=0$ in a consistent way independently of the
value of $D$. The answer to the first two questions is that
although the ND string, unlike the bosonic or supersymmetric
strings, does not sharply require a specific dimension such as 26
or 10, it nevertheless prefers $D$ to be as high as possible.

In relation to the first question, we note that the mass of the
scalar ground state is proportional to $D-1$; hence an infinite
$D$ is consistent with one end of the string being immovable.
Another consequence of this infinite $D$ limit is that the group
$SO(D-1)$ with $(D-1)(D-2)/2$ generators and the full inhomogenous
Lorentz group with only $2D-1$ additional generators "merge up to
order $1/D$" ; thus in a sense, the ND string recovers its
Poincar\'{e} invariance as $D$ goes to infinity. This is evidenced
also by a fact we observed earlier: the overestimation in the
density of states vanishes like $1/D$. Interestingly, In the
rather different approach to the problem taken in \cite{Kleinert},
$D$ has to be taken to infinity to make a saddle point calculation
possible.

The third question, namely the presence of the anomaly term in the
Virasoro algebra, gets affected by the above choice of $D$; the
coefficient of the anomaly is now infinite.  Since conformal
invariance is violated by the immobile end, a conformal anomaly
has to appear somewhere in the model; its becoming infinite is
perhaps an indication is that like most other infinite quantities
in physics, it is to be disregarded!  Indeed, the problems
normally associated with the conformal anomaly have already been
solved: there are no negative norm or zero norm states, and the
Virasoro conditions have organized and selected states into
$SO(D-1)$ irreps.

Finally, there is the intriguing question, which we intend to
investigate, of whether this system of half a bosonic string has
in some sense a fermionic character.  This is suggested not only
by the general fact that a pair fermions behaves like a boson, but
also by some specific hints.  Instead of restricting the range of
$\sigma$ to $\frac{\pi}{2}$ and working with odd oscillator modes,
one could keep the range as $\pi$ and work with half-odd integer
Neveu-Schwarz-like modes;  recall also the halving of the number
of dimensions appearing in front of the Virasoro anomaly noted
earlier.

Note added: After submitting the above note to the Web, we were
notified of two earlier papers \cite{Fair}, \cite {Sieg} involving
ideas related to ours.  Ref.\cite{Fair} is concerned with
off-shell states in Dual Resonance models; in particular, it
employs an $R^\mu$ field with Neveu-Schwarz-like half integer but
commuting modes for the factorization of amplitudes with one
off-shell particle. Had we chosen the interval $[o,\pi]$ instead
of $[o,\frac{\pi}{2}]$ for our $X^\mu$'s, we would have also
obtained half-integer commuting modes instead of our odd-integer
ones.  In \cite {Sieg}, the $D$ dimensions of spacetime are
separated into $D_o$ ordinary and $D_E$ extraordinary components
for the purpose of changing the intercept and the dimension to
physical values. The analysis is based on maintaining Poincar\'{e}
covariance in the ordinary dimensions while breaking it in the
others by fixing one end of the extraordinary $X^\mu$ coordinates.
In contrast, Poincar\'{e} covariance is entirely reduced to
$SO(D-1)$ covariance in our work.

\noindent {\bf Acknowledgements}
We are grateful to D. Fairlie and W. Siegel for bringing their
papers to our attention. We are grateful to S. Arapo\u{g}lu, M. Ar\i k, C. 
Deliduman, R.
G\"{u}ven and T. Turgut for useful discussions.


\newpage

\begin {thebibliography}{5}


\bibitem {Nambu}
Y. Nambu, "Quark model and the factorization of the Veneziano
amplitude", in \textit{Symmetries and Quark Models}, ed. R. Chand,
Gordon and Breach, p. 269 (1970); Y. Nambu, "Duality and
hadrodynamics", \textit{Lectures at the Copenhagen
Symposium}

\bibitem {Susskind}
L. Susskind, Phys.\textbf{ Rev.D1}
(1970)1182

\bibitem {Venez}
G. Veneziano, Nouvo Cim. \textbf{57A}(1968)190

\bibitem {DolenHS}
R. Dolen, D. Horn and C. Schmid, Phys. Rev. \textbf{166}
(1968)1768

\bibitem {Harari}
H. Harari, Phys. Rev. Lett. \textbf{22} (1969)562

\bibitem {Rosner}
J. Rosner, Phys. Rev. Lett. \textbf{22} (1969)689

\bibitem {Kleinert}
H. Kleinert, G. Lambiase and V.V. Nesterenko, Phys.LettB 384,
313-317, (1996)

\bibitem {Polch}
J. Polchinski, String Theory, Volume 1, p. 261, Cambridge
University Press (1998)

\bibitem {Polch'}
J. Polchinski, String Theory, Volume 1, p. 21, Cambridge
University Press (1998)

\bibitem {gsw3}
M.B. Green, J.H. Schwartz, E. Witten, Superstring Theory Vol.1.,
p. 116, Cambridge University Press, (1987)

\bibitem {gsw4}
M.B. Green, J.H. Schwartz, E. Witten, Superstring Theory Vol.1.,
p.118, Cambridge University Press, (1987)

\bibitem {Fair}
E. Corrigan and D. B. Fairlie, Nucl.Phys. B91 (1975)527

\bibitem {Sieg}
W. Siegel, Nucl.Phys. B109 (1976)244

\end {thebibliography}
\end{document}